\documentclass{egpubl}
 \JournalSubmission    % uncomment for submission to Computer Graphics Forum
\CGFccby
\usepackage[T1]{fontenc}
\usepackage{microtype}
\usepackage{textcomp}
\usepackage{dfadobe}  
\usepackage[notext]{stix} % to generate the correct dagger symbols
\usepackage{cuted}%appendix
\usepackage{cite}  % comment out for biblatex with backend=biber
\usepackage{booktabs}
\usepackage{tabu}
\BibtexOrBiblatex
\electronicVersion
\PrintedOrElectronic
\ifpdf \usepackage[pdftex]{graphicx} \pdfcompresslevel=9
\else \usepackage[dvips]{graphicx} \fi

\usepackage{egweblnk}
\makeatletter
\Wileypagenumber%
\copyrightTextTitPag{\p@journalSubmissionText{(\number\month/\number\year)}}
\copyrightTextRunPag{\p@journalSubmissionText{(\number\month/\number\year)}}
\def\ps@titlepage{\let\@mkboth\@gobbletwo
 \def\@oddhead{\raisebox{\z@}[8pt][1pt]{\parbox{\textwidth}{\small
   \hfill
 }}}%
 \def\@oddfoot{{\tiny\raisebox{\z@}[8pt][1pt]{\parbox[t]{18pc}{\sloppy
  \p@copyrightTextTitPag}}}\hfill}
 \let\@evenhead=\@oddhead
 \let\@evenfoot=\@oddfoot
 \let\sectionmark=\EmptySectionmark
 \let\subsectionmark=\EmptySubsectionmark
}
\renewcommand\p@journalSubmissionText[1]{%
~%submitted to COMPUTER GRAPHICS \textit{Forum} #1.%
}%
\makeatother

\usepackage{microtype}
\usepackage[svgnames]{xcolor}
\usepackage{svg}
\usepackage{enumitem}
\usepackage[capitalise,nameinlink,noabbrev]{cleveref}
\usepackage{comment} 
\usepackage{graphicx}
\usepackage{ccicons}

\title[A Design Space for Visualization Transitions of 3D Spatial Data in Hybrid AR-Desktop Environments]%
      {\vspace{-1em}A Design Space for Visualization Transitions of 3D Spatial Data \\ in Hybrid AR-Desktop Environments\vspace{-1.5em}}

\makeatletter
\def\orcid#1{\hbox{\href{\orcid@base #1}{\raisebox{.2em}{\includegraphics[height=.8em]{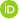}}}}}
\makeatother

\newlength{\authorsep}
 \author[Y. Lu et al.]
 {
 \parbox{\textwidth}{\centering Yucheng Lu (\raisebox{-.5pt}{\includegraphics[height=7pt]{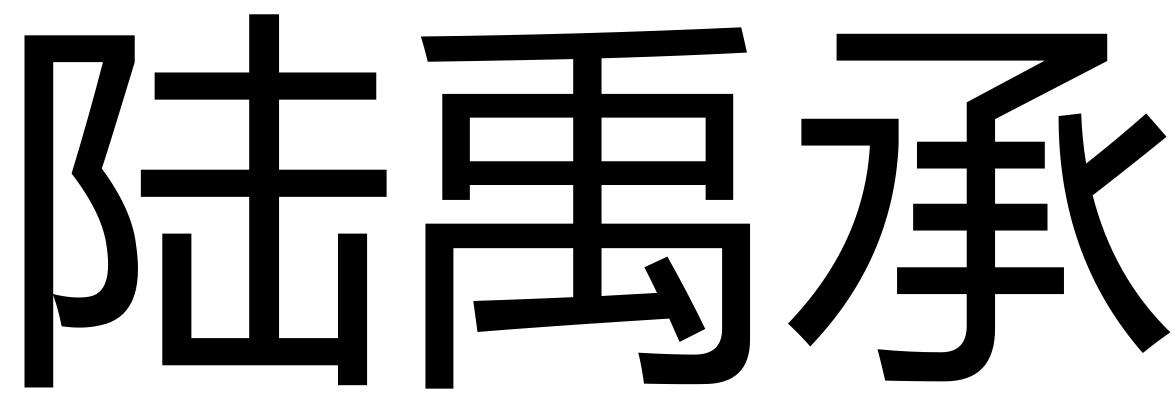}})\textsuperscript{1}\orcid{0000-0002-4131-6875}\hfill%
         Tobias Rau\textsuperscript{1,2}\orcid{0000-0002-3310-9163}\hfill%
         Benjamin Lee\textsuperscript{3}\orcid{0000-0002-1171-4741}\hfill%
         Andreas Köhn\textsuperscript{2}\orcid{0000-0002-0844-842X}\hfill%
         Michael Sedlmair\textsuperscript{2}\orcid{0000-0001-7048-9292}\hfill%
         Christian Sandor\textsuperscript{1}\orcid{0000-0002-3990-2728}\hfill%
         Tobias Isenberg\textsuperscript{1}\orcid{0000-0001-7953-8644}}
         \\
 {\parbox{\textwidth}{\centering $^1$ Université Paris-Saclay, CNRS, Inria, LISN, France\hspace{7mm}
          $^2$ University of Stuttgart, Germany\hspace{7mm}
         $^3$ JPMorganChase, USA%
        }
 }
 }
\newcommand{\changed}[1]{\textcolor{RoyalBlue}{#1}}
\renewcommand{\changed}[1]{#1}

\usepackage[normalem]{ulem}
\newcommand{\eg}{e.\,g.}
\newcommand{\ie}{i.\,e.}
\definecolor{softgray}{RGB}{200, 200, 200}
\newcommand{\ycr}[1]{\textcolor{softgray}{\sout{#1}}}
\newcommand{\rcr}[1]{\textcolor{gray}{\sout{#1}}}

\makeatletter
\renewcommand{\ycr}[1]{\@bsphack\@esphack}
\renewcommand{\rcr}[1]{\@bsphack\@esphack}
\makeatother

\newcommand{\osfrepo}{\href{https://osf.io/ve52m/?view_only=45718c6f028943078fe5f0ecf2668f8b}{\texttt{osf\discretionary{}{.}{.}io\discretionary{/}{}{/}ve52m}}}

\begin{document}
\teaser{%
%\vspace{-1.2em}% otherwise there is an empty line being generated
\vspace{-3em}
 \centering%
 \includegraphics[width=0.945\linewidth]{images/teaser.pdf}%
  \vspace{-0.5em}
  \caption{%
	Schematic views of XR transitions between 2D and 3D representations, applied to (A) a planar projection and its intermediate states, (B) a structure formula of a molecule and its 3D ball-and-stick representation, and (C) an MRI slice and its full volumetric representation.
  }
\label{fig:teaser}
}

\maketitle
%-------------------------------------------------------------------------
\begin{abstract}
%\todo{CGF require free format. Modify the journal name.}
%\todo{YC:Gray is removed text, blue is added text.}
We present a design space for animated transitions of the appearance of 3D spatial datasets in a hybrid Augmented Reality (AR)-desktop context. 
\changed{Such hybrid interfaces combine both traditional and immersive displays to facilitate the exploration of 2D and 3D data representations in the environment in which they are best displayed.
One key aspect is to introduce transitional animations that change between the different dimensionalities to illustrate the connection between the different representations and to reduce the potential cognitive load on the user. The specific transitions to be used depend on the type of data, the needs of the application domain, and other factors. We summarize these as a transition design space to simplify the decision-making process and provide inspiration for future designs.
First, we discuss 3D visualizations from a spatial perspective: a spatial encoding pipeline, where 3D data sampled from the physical world goes through various transformations, being mapped to visual representations, and then being integrated into a hybrid AR-desktop environment. The transition design then focuses on interpolating between two spatial encoding pipelines to provide a smooth experience.
To illustrate the use of our design space, we apply it to three case studies that focus on applications in astronomy, radiology, and chemistry; we then discuss lessons learned from these applications.}
%\textcolor{red}{TODO:}
   % The ABSTRACT is to be in fully-justified italicized text, 
   % between two horizontal lines,
   % in one-column format, 
   % below the author and affiliation information. 
   % Use the word ``Abstract'' as the title, in 9-point Times, boldface type, 
   % left-aligned to the text, initially capitalized. 
   % The abstract is to be in 9-point, single-spaced type.
   % The abstract may be up to 3 inches (7.62 cm) long. \\
   % Leave one blank line after the abstract, 
   % then add the subject categories according to the ACM Classification Index 
%-------------------------------------------------------------------------
%  ACM CCS 1998
%  (see https://www.acm.org/publications/computing-classification-system/1998)
% \begin{classification} % according to https://www.acm.org/publications/computing-classification-system/1998
% \CCScat{Computer Graphics}{I.3.3}{Picture/Image Generation}{Line and curve generation}
% \end{classification}
%-------------------------------------------------------------------------
%  ACM CCS 2012
   %(see https://www.acm.org/publications/class-2012)
%The tool at \url{http://dl.acm.org/ccs.cfm} can be used to generate
% CCS codes.
%Example:
\end{abstract}
%-------------------------------------------------------------------------
\section{Introduction}
%\todo{The gray text is the deleted old text. Other color is newly added text. Red is todo or comment text}
A myriad of domains and applications depend on 3D spatial data for their investigations.  
While 3D spatial data is presented in stereoscopic 3D views without loss of structural information, we need to resort to some form of projection when we display it on 2D screens such as computer monitors. 
The projections commonly used in computer graphics simulate a (pinhole camera) view of 3D spatial data, but they struggle with providing viewers with a proper depth impression, particularly in data visualization. 
Other projection methods and transformations can convert the 3D spatial data into a 2D visualization. 
A map projection~\cite{jenny_adaptive_2012}, \eg, facilitates an occlusion-free view of the surface of the Earth or, combined with other transformations, can give an overview of the surface of biomolecules~\cite{krone_molecular_2017}.
\changed{In other application scenarios, abstract 2D representations are used to increase readability and emphasize information such as topology.
Structural formulas of molecules, for instance, are closer to graphs than to projections.}
Suitable transformations are thus usually determined by the domain, the dataset, and the application. 
\changed{Such transformations, however, always bring drawbacks, primarily distortion and filtering information for the sake of readability.}
\changed{Being able to switch between a projection on a traditional 2D display and a stereoscopic view of the 3D dataset could combine the strengths of both (\eg, the ability to see the whole surface but also getting an undistorted sense of distances, angles, and directions).} 

\changed{Hybrid (mixed reality) environments \cite{feiner1991hybrid, Bornik2006Hybrid} allow users to interact with multiple systems that reside at different points on the rea\-li\-ty-vir\-tu\-a\-li\-ty continuum (RVC) \cite{milgram_augmented_1995} at the same time.
The combination of an immersive (\eg, AR, VR) and a non-im\-mer\-sive (\eg, desktop workstation, tablet) environment can display 3D and 2D visualizations in the environment that they were originally designed for: \eg, 2D views on a desktop alongside visualizations of 3D spatial data in AR.} 
\changed{Another benefit is that these hybrid systems support or even enhance established desktop workflows that involve visual data analysis and even data manipulation on 3D spatial data via 3D input and stereoscopic rendering \cite{cavallo_immersive_2019,frohler_survey_2022, rau_charpack_2024, rau_understanding_2024, liao2025seamlessvr}, without losing features of specialized desktop applications and the familiar input.}
\changed{The AR-Desk\-top hybrid system can thus be seen as providing the ``best of both worlds,'' while also extending the available workspace (\cref{fig:teaser}).} 
\changed{Visualizations in different manifestations of the RVC that the AR and desktop setup offers us (also called actualities \cite{auda_scoping_2023}), when simultaneously shown in a juxtaposed fashion, can provide a better data overview and improve the da\-ta-un\-der\-stan\-ding process \cite{wang_towards_2020}---com\-pared to a desk\-top-on\-ly setup.} 
\changed{The actual \emph{mental} transformation to connect a 2D with a 3D representation of the same data, however, is difficult due to the data's complexity and the differences of the representations, and the most beneficial 2D and 3D visual representations may differ substantially.}

% Why do we do transitions? 
\changed{While juxtaposing multiple 2D and 3D visualization mappings can compensate individual shortcomings, a smooth transition within a single visualization can provide a clearer connection between 2D and 3D representations. It is particularly beneficial when users wish to focus on interaction and intermediate transitional states, such as sliced volumes.}
\changed{One possible way to establish clearer connections between 2D and 3D representations in a single visualization is the use of animations \cite{TVERSKY2002247}.} \changed{In our case, they provide means for enhancing the mental connection between structural information in 2D and 3D and, consequently, reducing cognitive load.}
Studies indicate that animated transitions between different actualities can provide users with a clear picture of a given projection's distortion and can establish a connection between the two (or more) representations \cite{seraji_analyzing_2024}. 
Animations also improve people's understanding of the relationship between different visual mappings \cite{heer_animated_2007}.

% what are we doing? 
\changed{When exploring ways of mapping spatial datasets to 2D and 3D representations, placing and scaling these visualizations for an immersive and non-immersive environment, adding annotations, and animating the transition between the representations, the multitude of possibilities quickly becomes unmanageable. 
We thus propose a design space to better understand and manage possible considerations. 
While previous design spaces \cite{lee_design_2022,hong_survey_2024-1} have demonstrated the potential for representing abstract information in Mixed Reality, we focus on actuality-adaptive visualization transitions within hybrid environments, specifically for spatial datasets.}

%% How did we craft the design space? How is it grounded? 
\changed{To develop the design space, we first explored possible datasets and domains that provide data with high complexity in order to benefit from the proposed approach.
In addition, the datasets have to provide at least one 2D and one 3D representation that is well-known in their respective domains.
We then developed a prototype that implements the transition process of these meaningful representations---focusing on 2D and 3D visualizations of spatial datasets in a hybrid AR-desktop environment.
This prototype helped us to refine our design dimensions and to verify and showcase the design space with three example datasets (Exoplanet, Brain, Molecule).}

%% About the design space
\changed{In the first part of our design space we provide a description of the initial and the target visualization, which consists of three main parts: (1) the \emph{visual mapping} that describes transformation and the resulting dimensionality of the representation, (2) \emph{annotations} that provide supporting indicators with the goal of increase readability, and (3) \emph{pose in hybrid environment} that defines where the visualziation is located and how it is scaled.}
\changed{In the second part, we describe possible design dimensions when animating the interpolation between the initial and the target visualization.}
\changed{With these elements we can describe a variety of visualization presentation setups in the AR-desktop environments and apply seamless \emph{transitions} between different visualization setups with interpolation and animation.}
In addition, we further discuss considerations that focus on adopting the design space to a wide range of 3D spatial data (\cref{sec:discussion}). 
With the design space and the discussion of high-le\-vel design considerations, we aim to inspire the creation process of hybrid systems with animated actuality adaptive visualization transitions. 

\changed{To illustrate the utility of our design space, we use our prototype to demonstrate this system as well as our design space with three case studies (\autoref{sec:case_studies_sec}): (i) astrophysics data with star systems that have exoplanets; (ii) magnetic resonance imaging (MRI) scan of a human brain; and (iii) molecular structures of different complexity.}

In summary, we contribute a design space for 3D spatial data visualization transitions and lessons learned from three case studies with datasets from astrophysics, medical visualization, and chemistry. %that illustrate the design space.

\section{Related work}
Our work describes transitions of visualizations that change their dimensionality corresponding to the environment. 
\changed{In particular, we work with a hybrid system that uses AR and desktop PC environments and relies on projection and animation, as we discuss next.}

\subsection{Cross-reality systems}
Many approaches can facilitate a seamless blend of different actualities, which have been surveyed by Auda et al.~\cite{auda_scoping_2023}. 
\changed{For example, users can \emph{transition} from one actuality to another, which is also known as a transitional interface (TI)~\cite{billinghurst_magicbook_2001}.}
\changed{From the three types of systems that Auda et al.\ derived, however, the \emph{substitutional} one is the most relevant for us as we focus on scenarios in which a user controls virtual objects that transition between different environments \cite{lee_design_2022, lee_deimos_2023, seraji_analyzing_2024, Rau:2025:TDR}.}
\changed{Fr{\"o}hler et al. \cite{frohler_survey_2022}, \eg, use \emph{cross-virtuality} to describe visualization applications that transition between different actualities, and Zagermann et al. \cite{zagermann2022complementary} think of cross-device interaction, like ours, as \emph{complementary interfaces}---in our case, the AR and the desktop complement each other to take advantage of both 2D and 3D spaces.}

Existing literature mainly focuses on interaction design. 
A design space of single users interacting with a hybrid system was proposed by Wang and Maurer~\cite{wang_design_2022}.
In one of their scenarios, a user moves a visualization from one point of the RVC to another, which involves a transformation of the visualization from 2D to 3D. 
A similar scenario was implemented and studied by Schwajda et al.~\cite{schwajda_transforming_2023} using graph data that transitions from a large-scale display into an AR environment. Another implementation by \changed{McDade et al.~\cite{mcdade_realitydrop_2023} renders a 3D model as an exploded view on the PC and as a merged model in AR.}
\changed{While these approaches~\cite{gall2022cross} focus on the aspects for supporting users to form a joint mental model of both visualizations, Lee et al.~\cite{lee_design_2022} formulated a general design space for systems that incorporate abstract data visualization transitions from 2D to 3D and vice versa in a single actuality.}
\changed{In our work we focus on spatial data visualizations that the design space of Lee et al.~\cite{lee_design_2022} did not consider or could not support.}

\changed{We also note that, in a AR-desktop hybrid environment, augmenting 2D content in desktop monitors with 3D AR content is a common approach. Gall et al. \cite{gall2023uncertainty}, \eg, enhance the uncertainty visualization of the distribution by providing extra 3D AR visualization and facilitating gestural interaction. They and others apply this approach to tomography visualization \cite{gall2022cross, gall2024immersive, morthmixed}. Other studies focus on the interaction techniques for transferring virtual objects from desktop monitors into the AR space.  Cools et al. \cite{cools2022towards} propose a framework to expand the desktop monitor to AR space through an arced virtual screen. Rau et al. \cite{Rau:2025:TDR} explore various forms of gestural interaction that enable users to bring desktop monitor content into AR space. All of these works also show that an AR-desktop hybrid \ycr{can provide extra engagement and} can complement the shortcomings of traditional desktop visualization, yet the spatial visualization transition design is still unclear---which is what we add.}\vspace{-.75pt}

\subsection{Projections and mappings}
For further discussion of the related work and, later, the design space we need to define three terms. 
First, \emph{transformation} is an operation, \eg, a projection or an algorithm, that maps a visualization between 2D and 3D. 
Second, \emph{transition} is the process of changing the actuality of a visualization. 
In our specific case, we change the visualization's environment from 2D to 3D or from 3D to 2D. 
Third, \emph{morph} is the continuously animated shift between visualization representations. 
\changed{In general, a morph is independent of the environment in which the visualization is rendered, but for our proposed design space we consider that a transition always incorporates a morph.}

Visualizations of 3D spatial data or descriptions based on the 3D physical world~\cite{card_readings_1999} facilitate the recall of the physical spatial understanding. 
Nowadays, 3D visualizations are still predominantly rendered and analyzed on traditional PC workstation setups that project the final image on 2D displays~\cite{hurter_fiberclay_2019}. 
Visualizing 3D datasets on 2D screens, however, suffers from occlusion~\cite{lawonn_occlusion-free_2016}, and perspective distortion~\cite{petkov_interactive_2012}. 
For a good spatial understanding of a perspective view of complex 3D visualizations, users have to navigate the scene~\cite{dix_starting_1998}, which can be difficult using standard workstation input modalities. 
Hurter et al.~\cite{hurter_fiberclay_2019} directly compared 3D visualizations projected on a 2D screen with the immersive space, and suggested that the immersive visualization ``fosters the discovery of many additional insights.''

2D projections of 3D data, in contrast, can be more comprehensible than 3D visualizations \cite{neri_time_2006, borkin_evaluation_2011-2}. 
Geometric and abstract projections are commonly used to simplify the depiction of 3D spatial data, such as Earth map projection methods in geography \cite{snyder_map_1987, lisle2004stereographic, jenny_adaptive_2012}. 
\changed{While distortion is inevitable due to non-isometric transformations, projections retain specific visual characteristics. The Azimuthal Equidistant projection \cite{anderson_oblique_1974}, \eg, keeps the distance undistorted when \mbox{projecting} the globe onto a plane.}
\changed{In medicine, conformal mapping is widely used to locally preserve small visual shapes by keeping \mbox{internal} angles invariant \cite{kreiser_survey_2018}.} 
\changed{Physical 2D screens are considered precise and fast for 2D content, but limiting for 3D content~\cite{Bach2018displays}.}
\changed{Although it is debated whether 2D outperforms 3D or vice versa~\cite{Tory2006performance,Horak_2019_Vistribute,lu_effect_2023}, hybrid interfaces are a valuable option that benefits from both, 2D and 3D~\cite{Dubel2014_2d3d, hong_survey_2024-1}.} 
\changed{Users tend to make use of both 2D and 3D visualizations to complete their tasks depending on the circumstances \cite{neri_time_2006}, thus combining their advantages \cite{meuschke_combined_2017}.} 
\changed{Certain 2D and 3D visual representations are inherently more compatible with the affordances of 2D screens and 3D AR environments, and are thus particularly suitable for hybrid settings.}
\changed{Building on these foundations, we develop our hybrid AR–desktop environment and the design space.}\vspace{-.25pt}
 
\subsection{Animation and transition}

\changed{Animation transitions have been widely adopted in various applications, often to facilitate spatial understanding during transitions between Augmented Reality (AR) and Virtual Reality (VR)~\cite{pointecker2024real,liao2025seamlessvr} or to help users track changes between visualizations \cite{elmqvist2008roll}.
Liao et al. \cite{liao2025seamlessvr}, \eg, proposed an approach to smoothly morph 3D models from a 2D to a 3D representation between AR space and the monitor. They show that the animated transitions enhance user engagement and support perceptual tracking of objects across different representations.
In our work, the transitions occur between AR and desktop environments, accompanied by changes in the visualization itself. We focus on transitions that adapt to the AR space or physical monitor displays as well as on animation designs used to handle changes in visual encoding.}

\changed{To design the animation, several common designs already exist, such as animation curves \cite{dragicevic_temporal_2011}, staggered animations, or approaches for controlling animations \cite{abukhodair_does_2013}. We include some of these established designs and further add an additional interpolation method dimension for the design of animations in transitions.}

\section{Design Considerations}
\label{sec:design_space}
%
% figure is located here:
% https://www.figma.com/design/8AnZayOpClzuXGUWkGBo7a/AR-Transition?node-id=174-623&node-type=frame&t=1JI3ULothUnlvYDz-0
%
\changed{Below, we begin by briefly summarizing our overall process for arriving at the design space, including our core considerations for 2D and 3D representations of spatial visualizations and the resulting design of transition animations.}\vspace{-.55pt}

\subsection{Development process}
Spatial data presents inherent positional information that necessitates careful consideration of how to encode it into visual representations. This positional information is also essential in the design of animations, particularly when transitioning between 2D and 3D visualizations. To develop our proposed design space, we centered the transition design around the concept of spatial information.

\changed{We started by reviewing relevant literature on foundational concepts and looked for datasets and domains that are well-suited as examples and potential applications of the design space.}
\changed{We found that datasets in the domains of medicine, geography, and chemistry cover a wide range of scenarios that we want to support with our design space.
Each dataset has an inherent 3D representation and at least one unique 2D mapping (here: projection, slice, and graph representation) that is wildly different from the other data domains and their respective 3D representation.}
Further, prior research has summarized design choices for 2D and 3D visualizations~\cite{hong_survey_2024-1}, design space for transitioning non-spatial data~\cite{lee_design_2022}, and various geometry projection techniques for 2D and 3D visualizations, such as flattening or projection techniques~\cite{kreiser_survey_2018, krone_molecular_2017, Karen2001Symbolization, snyder_map_1987}.
\changed{The chosen domains and studies encompass a wide range of data types, including volumetric data, point-based data, and topological data, which are frequently encountered in various scientific and engineering domains.}
Next, we examined existing systems that design and implement transitions within XR environments~\cite{hurter_fiberclay_2019, miao_dimsum_2018, halladjian_scale_2020, mohammed_abstractocyte_2018, yang_maps_2018, schwajda_transforming_2023}. 

\changed{While Lee et al.~\cite{lee_design_2022} explored design choices for non-spatial datasets and interaction, our interest lies in how 3D data is expressed based on its inherent spatial characteristics and how it is maintained during the transition, within the hybrid AR-desktop environment. To realize this goal, we extend their existing visualization design, but deliberately omit specific design dimensions such as input modalities or other devices.}
We focus on the transition relationship between the physical screen and AR space and thus generalize our design space based on the exploration of the three mentioned datasets.\vspace{-.55pt}

\subsection{\changed{Intermediate Visualization}}

\changed{Transitions in visualization involve changes in specific design dimensions. 
A common scenario in AR-desktop environments is to transition from a 3D model to a flattened 2D view by altering the positional encoding \cite{kreiser_survey_2018}. 
Yet each individual visualization is produced through a set of design decisions, so transitions primarily reflect a shift between two distinct visualizations or two sets of design choices. 
We thus focus on designing these transitions to manage the changes between two sets of design choices.} 

\changed{Transitions between design choices, however, may also produce intermediate visual representations, which result naturally from the interpolation during transitions between the two sets of designs. Many, if not most, of these intermediate visual representations only facilitate a visually smooth animation but are semantically insignificant: they do not necessarily represent an actual state of the data as they show a distorted intermediate state, which likely reduces the visualization's readability. In certain cases, however, the intermediate representations can be meaningful if they shift the presentation to reveal spatial data from a different perspective (as we later, in \autoref{sec:case_studies_sec}, demonstrate in our case study on volumetric data). Nonetheless, the specific sequence and animation of the intermediate designs can have a substantial impact on the effectiveness of the transition.}

\section{Design Space}
    
\begin{figure*}[t]
  \centering
  \includegraphics[width=\textwidth]{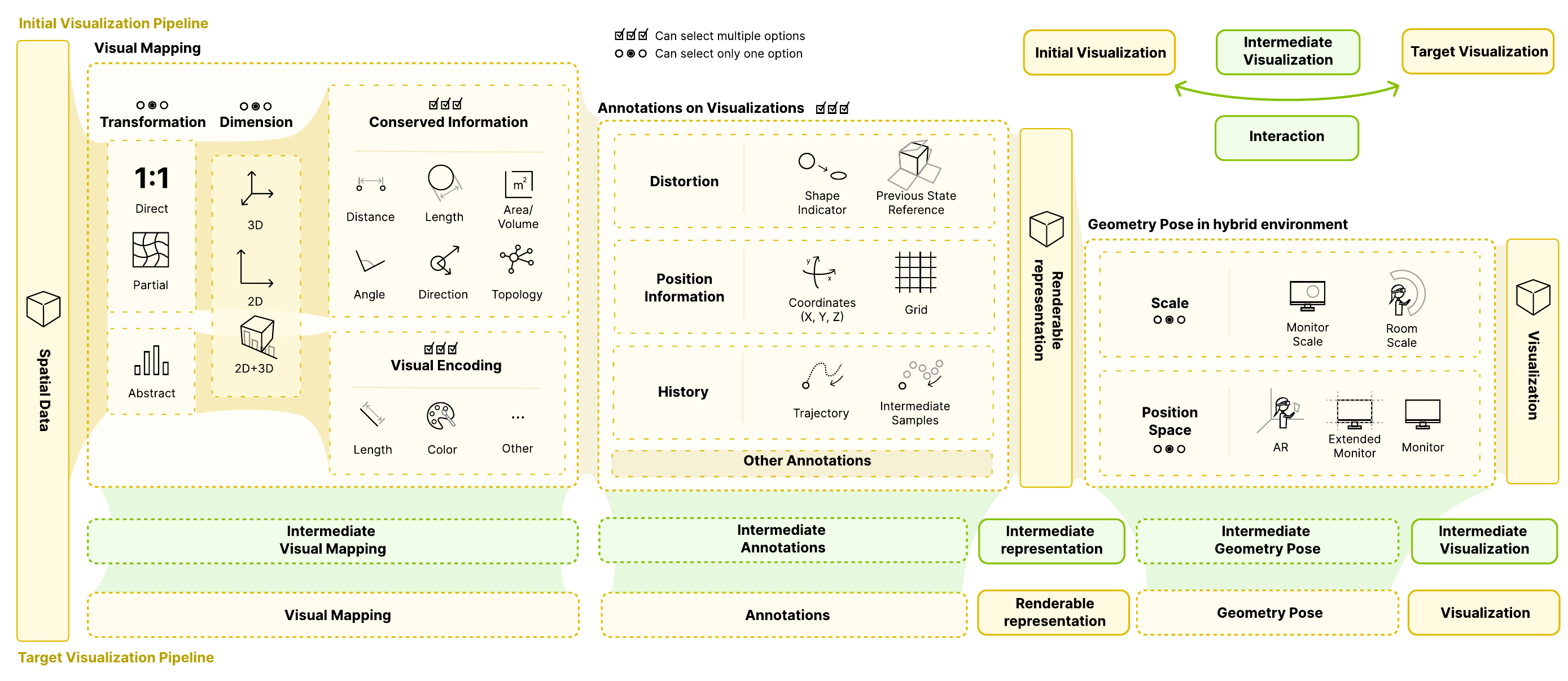}\vspace{-5pt}
  \caption{%
  %\todo{the caption is replaced.}
  \changed{Our design space for the visualization pipeline consists of three stages: The visual mapping, the additional annotations, and the geometry pose of the visualization embedded within the AR-desktop space. From left to right, the designer can make the design choices at each stage and create the final visualization. After several visualizations come out through this pipeline with slightly different design choices, we can determine the design for transitioning between them, as we illustrate in \autoref{fig:design-space-transition}.}}\vspace{-4pt}%
  \label{fig:design-space}
\end{figure*}

\changed{Based on these considerations, we separate our transition design into two parts: (1) the design of the initial and the finial visualization (\cref{fig:design-space}) and (2) the design of the transition including intermediate states between two sets of design choices (\cref{fig:design-space-transition}).}
The parameters we discuss within these two parts describe possible degrees of freedom for variation of the specific transition, as we describe next.

\subsection{Pipeline}

We use a pipeline to describe how a visualization is presented, and therefore, how the spatial data undergoes several design choices before finally being presented as a visualization (\cref{fig:design-space}).
\changed{Specifically, the \textbf{Visual Mapping} maps the data to the renderable visualization (Bruckener et al.'s \cite{Bruckner:2019:MSD} \emph{visualization mapping} or a common visualization pipeline \cite{card_readings_1999})}. The \textbf{Annotation} adds extra information to enhance the spatial understanding and the \textbf{Geometry Pose} discusses putting the renderable visualization in the actual environment (Bruckener et al.'s \emph{output mapping}).

\subsubsection{Visual mapping}
\label{ssec:visual_mapping}

The visual mapping is the most important step of the visualization pipeline that consists of a set of transformations, which ``repeatedly transform data into different forms and ultimately transform it into a representation that can be rendered by the computer system''~\cite{schroeder_overview_2005}. 
This set of \textbf{transformations} may retain some of the intrinsic properties of 3D spatial datasets:
geometry, topology, and attributes~\cite{schroeder_overview_2005}. 
\textbf{Direct} transformations do not alter any of the intrinsic properties of the dataset. A molecule's balls-and-sticks representation, \eg, does not change its properties throughout the transition. 
\textbf{Partial} transformations alter some but not all of the intrinsic properties and possibly alter the dimensionality of the representation. 
One example of a partial transformation is a map projection, which alters geometry and topology, but not necessarily attributes, \eg, landmass or elevation. \changed{Another example is to use a subset of the dataset, such as a slice of the volumetric data, or to use transparency \cite{waldin2016chameleon} to hide a part of data and features.}
An \textbf{abstract} transformation does not retain any of the intrinsic properties of the spatial dataset, \eg, a plot of the total energy of a system. \changed{For example, Assor et al.~\cite{assor2024exploring} transition between abstract bar chart presentation and concrete visualization that includes intrinsic physical properties.}
Another example is mapping the count of different proteins in a virus to a bar chart~\cite{sorger_metamorphers_2017}.

The \textbf{dimension} impacts the way we display the visualization; it also later impacts the visual encoding considerations. \changed{A 2D representation may inherently partially discard positional information from the spatial dataset during the visual encoding.}
Depending on the chosen transformations, we consider the dimensionality of the resulting renderable data to be one of the following categories:
\begin{itemize}[nosep,topsep=-1ex]
    \item \textbf{2D} visualizations are well suited for desktop PCs with a physical monitor setup. If the original data, however, is hi\-gher-di\-men\-sio\-nal (3D or nD), we first have to project it to two dimensions, such as via dimension reduction~\cite{vernier_quantitative_2020}, losing some of the original data information.
    This loss of information can (partially) be compensated for by extra annotations or through interaction.
    \item \textbf{3D} visualizations, in particular those of 3D spatial datasets, are naturally compatible with the AR environment. For 2D representations, extra information can be encoded on the extra dimension of the 3D representation. 
    Time-based data can represent a specific slice of information at a given moment in 3D~\cite{kwan2000highPerformance}, although this approach does not capture the entire dataset. 
    \item \textbf{Hybrid 2D and 3D} combinations have the potential to reveal additional results~\cite{hong_survey_2024-1}. Therefore, such hybrid representations are well-suited for the AR-desktop hybrid environment. Ideally, the 2D part of the visualization is displayed on the monitor, and the 3D part in AR in a juxtaposed fashion~\cite{schwajda_transforming_2023}.
\end{itemize}

It is important to determine which spatial information should be preserved (\textbf{conserved information}) to select the most appropriate visual mapping for a given dataset and application. 
For the actual implementation, it is also possible to pick several categories and mix them to retain part of each spatial attribute~\cite{nadeem2017spherical} or add customized constraints like parallel or self-in\-ter\-sec\-tion~\cite{merry2006animation}. 
A transformation that conserves the specified properties, however, does not necessarily exist for every possible dataset.
\begin{itemize}[nosep,topsep=-1ex]
    \item \textbf{Distance, length} provides invariance of distance (isometry)---one of the most important geometric properties to provide accurate measurements between two points. While isometry is hard to conserve within complex datasets with curved surfaces, it is still worth keeping the distance distortion to a minimum.
    \item \textbf{Area, volume} are commonly used in geographic maps~\cite{snyder_map_1987} as area-equal projections. When the distance is inevitably distorted, the invariance of the area will conserve scale information.
    \item \textbf{Angle, direction} mostly refer to conformal map\-ping~\cite{petkov_interactive_2012}, the visualization will keep its shape during a transition.
    \item \textbf{Topology} is commonly modified if the visualization is separated into several parts like in an exploding view \cite{bruckner2006exploded} or a 3D model is expanded to a surface \cite{kreiser_survey_2018, krone_molecular_2017}. The loss of topology will impact the global structure information but potentially benefit local exploration. 
\end{itemize}
\textbf{Visual encodings} are commonly used to represent data using visual variables by adjusting visual marks. 
For inherently spatial data, however, we need to be cautious about maintaining its spatial information. 
Visual encoding usually has to---at least partially---retain spatial information when, \eg, the original data is projected to a lower dimension. 
\changed{Thus, 3D or 2D spatial data is, most commonly, directly mapped using its positional information, with other visual channels such as color being used for additional data layers.}
This approach retains the original data semantics and aligns with the users' spatial and arrangement intuition. 
It also works well with AR environments due to the addition of stereoscopic vision and 3D input methods. 
If, in contrast, the \textbf{1:1} spatial mapping is replaced with \textbf{partially} spatial or with a non-spatial \textbf{abstract} one, such as to avoid occlusion or to have a quick overview by means of projection, distortion becomes an issue. \changed{The transition from a 1:1 globe representation to the Mercator projection, for instance, as illustrated in \autoref{fig:case_study_figures}(C), significantly distorts areas near the poles, but also provides a quick overview.}
The same effect also occurs when a user ``returns'' a visualization from the 3D AR space to the 2D monitor space due to the latter space's lower dimensionality.

\subsubsection{Geometry Pose}
The \textbf{Geometry Pose} describes the scale and position of the visualization relative to the available scales and to the anchor positions in the AR-desktop environment setup. 
The first of its aspects, the \textbf{Scale}, impacts how the visualization is embedded in the environment~\cite{yang_maps_2018}. 
While previous work~\cite{Cheliotis2021Classification} defined the scale based on the human size, we are (like others \cite{Zhao:2024:MTC}) primarily concerned about the interaction with the visualization within the AR-desktop hybrid environment and thus classify it according to the visualization's size relationship as it is embedded in a space: 
\begin{itemize}[nosep,topsep=-1ex]
    \item \textbf{Monitor scale} describes a small scale of visualization that can be embedded anywhere, like users' hands (hand scale), physical monitors, or a desk (desk scale). This scale is the easiest to integrate into our proposed AR-desktop environment. 
    \item \textbf{Room scale} is large and cooperates well with AR space and creates the opportunity to inspect parts of the visualization in detail or get a completely new perspective (\eg, an inside-out view~\cite{yang_maps_2018}). In desktop settings, the monitor can no longer display the whole visualization due to its constrained screen size. 
\end{itemize}

In our case, a visualization's current position depends on the device that is displaying it, as well as its current visual mapping:
\begin{itemize}[nosep,topsep=-1ex]
    \item The \textbf{AR space} renders a visualization usually stereoscopically and in 3D, but 2D visualizations are also acceptable.
    \item The \textbf{monitor space} is usually constrained by a rectangular plane, although curved or alternatively shaped monitors are included. This space best suits 2D visualizations, while 3D visualizations can be displayed using, \eg, a perspective projection.
    \item An \textbf{extended monitor space} reduces the constraint of the monitor. Therefore, it is possible to virtually extend the monitor to an almost endless plane or surface.
\end{itemize}

\subsubsection{Annotations}
Additional information provided by annotations can help viewers to understand the data, \eg, by giving visual indicators of the current distortion in a static view or during a \ycr{state} transition. 
Many forms of annotations are possible based on the design intention to further support users in connecting visual mappings during the transition by providing different visual cues. \cref{fig:case_study_figures}(D) shows several example annotations that illustrate their usage, which we discuss in more detail in our exoplanet case study in \cref{sec:case_studies_sec}.
\changed{Below, we focus on the annotations that visualization designers annotate structural information, since it is the most crucial part for spatial data. Other common annotations, such as labels, can, of course, be used as well.}
\begin{itemize}[nosep,topsep=-1ex]
    \item \textbf{Distortion cues} are intended to address the issue that distortion is difficult to detect for a given visualization without visual aids, especially for common map projection techniques. A \textbf{shape indicator}, such as the Tissot indicator~\cite{Karen2001Symbolization}, can directly show the effects of distortion when it is compared to the original physical spatial meaning. A duplicate view of a \textbf{previous state reference} also enhances distortion awareness. 
    \item \textbf{Position information} provides a relative reference for improving the locating quality because the direct visual mapping of position is likely inadequate to convey the visualization information.  \textbf{Coordinate} axes and coordinate \textbf{grids} are common means to improve readability of positional mappings. 
    \item A \textbf{history} can enable users to keep track of position information during the transition\, similar to motion lines \cite{Joshi:2005:IIT,Schmid:2010:PME}. Visualizing a \textbf{trajectory} is a common way to track a sequence of data samples. 
    \textbf{Intermediate samples} is a similar method for tracking the transition in which a full visualization is displayed. It can be applied when a trajectory is not feasible.
\end{itemize}

The mentioned annotation methods provide a general (but not exhaustive) view of possibilities in this category. 
Annotations usually cooperate well with partial transformations of the 3D spatial data and thus help the user to be aware of distortion and avoid implicit bias from distorted visual properties like length or size \cite{Ceja2021Bias}. 
Without a direct way to show the connection between the states, such as an animated transition (morph) between the visualizations, however, it remains mentally demanding to make the connection between the two visualizations. 

\subsection{Transitions}

\begin{figure}[t]
  \centering
  \includegraphics[width=\linewidth]{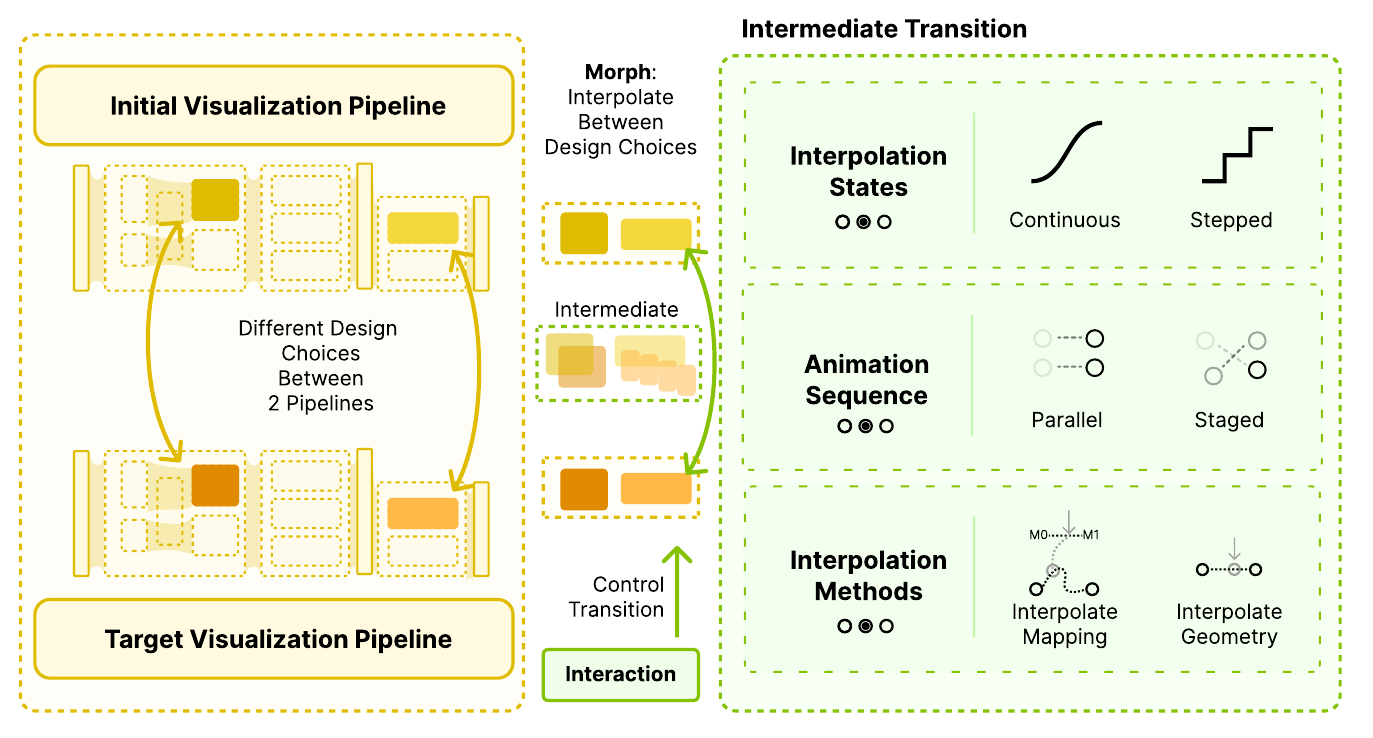}\vspace{-3pt}
  \caption{%
  \changed{The transition between two visualizations is primarily determined by the differences in design choices between the two visualization pipelines. For each distinct pair of design choices, designers can choose intermediate transition design options (right side) to define how to transition between them. The progression of these transitions is controlled through user interaction.}}\vspace{-2pt}
  \label{fig:design-space-transition}
\end{figure}

\changed{Once the design choices for the initial and target visualization have been made (or were given by an application domain), we need to specify the morph between them.}
For each morph, we focus on three factors: (1) the states of the interpolation, (2) the morph order of all visualization glyphs or elements, and (3) the interpolation method (see \autoref{fig:design-space-transition}). 
A common way to carry out a morph is to employ it as an animation. 
A well-de\-signed animation can provide a smooth morph experience and result in a higher engagement and understanding for the user~\cite{heer_animated_2007}. 
Still, it can also become time-con\-su\-ming to need to wait until the animation ends \cite{TVERSKY2002247}. 
The morph, however, can also be used to explore intermediate states between the initial and final ones---if they are meaningful. \changed{Such a meaningful visualization, however, is usually carefully designed, and a direct interpolation of geometry toward or away from it can easily break the envisioned representation and become confusing.}

The \textbf{interpolation states} is best designed based on the visual mapping. 
Many transformations are hard to understand or to explain well, even with adequate annotations. 
Separating animations that would normally happen simultaneously can provide substantial benefits for people to more easily understand the connection between the two states and sub-states~\cite{bederson_does_1999}.
\begin{itemize}[nosep,topsep=-1ex]
    \item \textbf{Continuous} interpolations can provide viewers with a smooth morph of the visualization, which can be used for simple transformations that do not cause severe self-interaction or occlusion during the transforming procedure.
    \item \textbf{Stepped} interpolations have one or several independent steps during the animation. This method facilitates the separation of changes that otherwise occur simultaneously in the morph: it puts these changes into a sequential order. In the Mercator projection, \eg, the first step is to project the data to a cylinder, and the second step is to unfold the cylinder into a rectangle~\cite{snyder_map_1987}.
    We can also combine multiple states and present them as a cohesive stepped animation (\eg, \autoref{fig:molecules}(d, e), where the animation consists of two steps: position transition and symbol transition).
\end{itemize}

The \textbf{animation sequence} focuses on how to animate each glyph or element of the visualization in order. 
Unlike the \textbf{interpolation states}, the animation sequence focuses on how to animate the whole collection of glyphs or elements in a specific time order. 
\begin{itemize}[nosep,topsep=-1ex]
    \item A \textbf{parallel} sequence transforms the whole visualization at the same time. It is common to use this mode for simple projections and data with meaningful intermediate states.
    \item A \textbf{staged} sequence starts with one or a few glyphs or elements and subsequently animates others. This approach can be responsive to user interaction, providing means to explore intermediate states interactively. For example, when the user ``returns'' a visualization and ``pushes'' it onto the PC screen, only those (3D) glyphs or elements of the visualization that begin to touch the screen surface start their animation procedure.
\end{itemize}

A straightforward interpolation of the element or glyph positions between the initial and target visualizations is possible for simple projections, but might disrupt the conserved information of the original visual mapping.  To preserve the constraints of the visualization and the semantic connection between initial and final visualizations, the interpolation method may account for these factors and should be implemented with alternative algorithms. 
Ultimately, two major forms of \textbf{interpolation methods} are possible:
\begin{itemize}[nosep,topsep=-1ex]
    \item The \textbf{interpolation of geometry} is a common and simple way to morph between two states by directly interpolating the final geometry positions of the initial and final visualizations. For each vertex of a geometry, its initial and final position is computed, and then the designed interpolation function between these two positions is applied. This method may distort the conversed information for the initial and target state, such as angles and area.
    \item The \textbf{interpolation of mapping} transitions between two visual mappings at the conceptual level, including the \textbf{conserved information} we had mentioned before. For instance, if the initial visualization conserves area while the final visualization conserves angles, the interpolation between both mappings will gradually reduce the preserved area concept and increase the preserved angle information concept of the visualization. When both visualizations share the same property, the mapping interpolation naturally keeps this property during the morph, offering an ideal continuum~\cite{petkov_interactive_2012} within the intermediate states that potentially provide a meaningful representation. Another simple example is the unfolding of a cube. Linear interpolation would distort the edges of the cube, while the interpolation of the mapping constrains the length of the edges. 
    Except for conserved information, we can also interpolate between different visual encodings, for which we show an example in \autoref{fig:molecules}. Here, the transition switches between the molecule glyph and chemical element symbols.
\end{itemize}

\section{Case Studies}
\label{sec:case_studies_sec}

To illustrate the utility of our design space, we now discuss three case studies: an exoplanet dataset, medical MRI data, and chemical molecule data. 
\changed{We believe that these case studies from different application domains of scientific data analysis showcase a representative amount of design choices possible in our design space.}
\ycr{For this purpose, we implemented the characteristics of our design space using Unity and the Mixed-Reality Toolkit (MRTK). 
We deployed the application on the Oculus Quest 3 and HoloLens 2 headsets for AR and a Windows desktop to build an AR-desktop hybrid environment. 
We synchronize the virtual object transformation, state, and user interaction behavior between the two devices through a local Wifi network connection. 
In this prototypical implementation, we realized various state conditions and transitions we discussed, with which we can now showcase a seamless screen extension that allows the user to drag a visualization off the screen using the mouse or gesture input.}
We refer the reader to the accompanying video for a visual explanation.

\subsection{Implementation}
\changed{To implement the prototypes, we used Unity and the Mixed-Reality Toolkit (MRTK) to develop two different applications targeting the desktop platform (Windows) and AR headsets. We deployed the AR application on either the Oculus Quest~3 or the Ho\-lo\-Lens~2, and ran the desktop component on a Windows desktop, together forming our AR-desktop environment.}
\changed{To set up this environment, we first calibrated the physical monitor by pointing to three of its corners in AR space with the index finger. We also established a local Wi-Fi network to synchronize user input and virtual object data between the AR and desktop platforms in real time.}

\changed{Our prototype incorporates a basic interaction mechanism that facilitates the transition across devices. Users can drag visualizations from desktop monitors into the AR space and vice versa. This drag operation can be performed using either hand gestures or a mouse. A typical interaction involves dragging a visualization off the desktop screen, triggering a transition from 2D to 3D visualization.}
\changed{Based on this implementation, we now describe the three case studies to illustrate our design space and showcase its descriptive power.}

\subsection{Exoplanet dataset}
% Background
As a representative of sparse 3D point data, we selected NASA's dataset of stellar systems with known exoplanets \cite{NASA:Exoplanets}, which records information about the respective stellar system and its planets. To describe the location of a system, experts use the International Celestial Reference System (ICRS)~\cite{ICRF_1998}---a spherical coordinate similar to a geographic map. 
It is thus meaningful to use projection techniques similar to a terrestrial map, which in turn can be represented as a 3D sphere or projected onto a 2D plane using various established projection techniques. Here, we visualize primarily the locations of the respective stars, as opposed to the information about the planets (which are naturally much closer to their home star than the stars are separated from each other on average).

\begin{figure}[t]
  \centering
  \includegraphics[width=\linewidth]{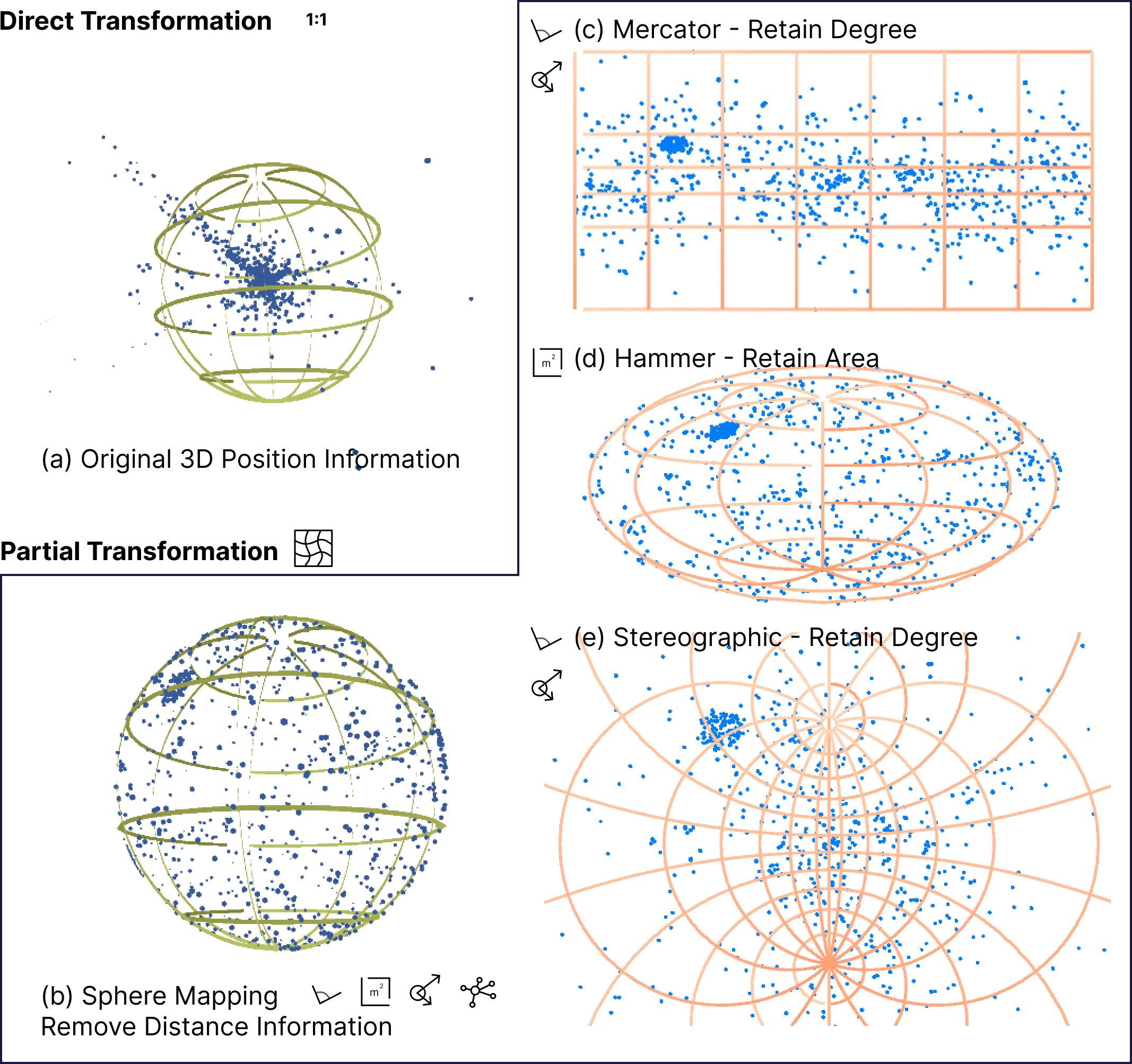}\vspace{-2pt}%{images/figure_stellar_encoding.pdf}
  \caption{%
  Visual mapping examples for the exoplanet dataset. 
  3D: (a) direct mapping of the dataset with ICRS spherical coordinates as annotation; (b) location of the exoplanets orthogonally projected onto the ICRS surface. 2D: (c) Mercator, (d) Hammer, and (e) stereographic projection. All projections use a grid as an annotation.}
  \label{fig:stellar_encoding}
\end{figure}

\begin{figure*}[t]
  \centering
  \includegraphics[width=\textwidth]{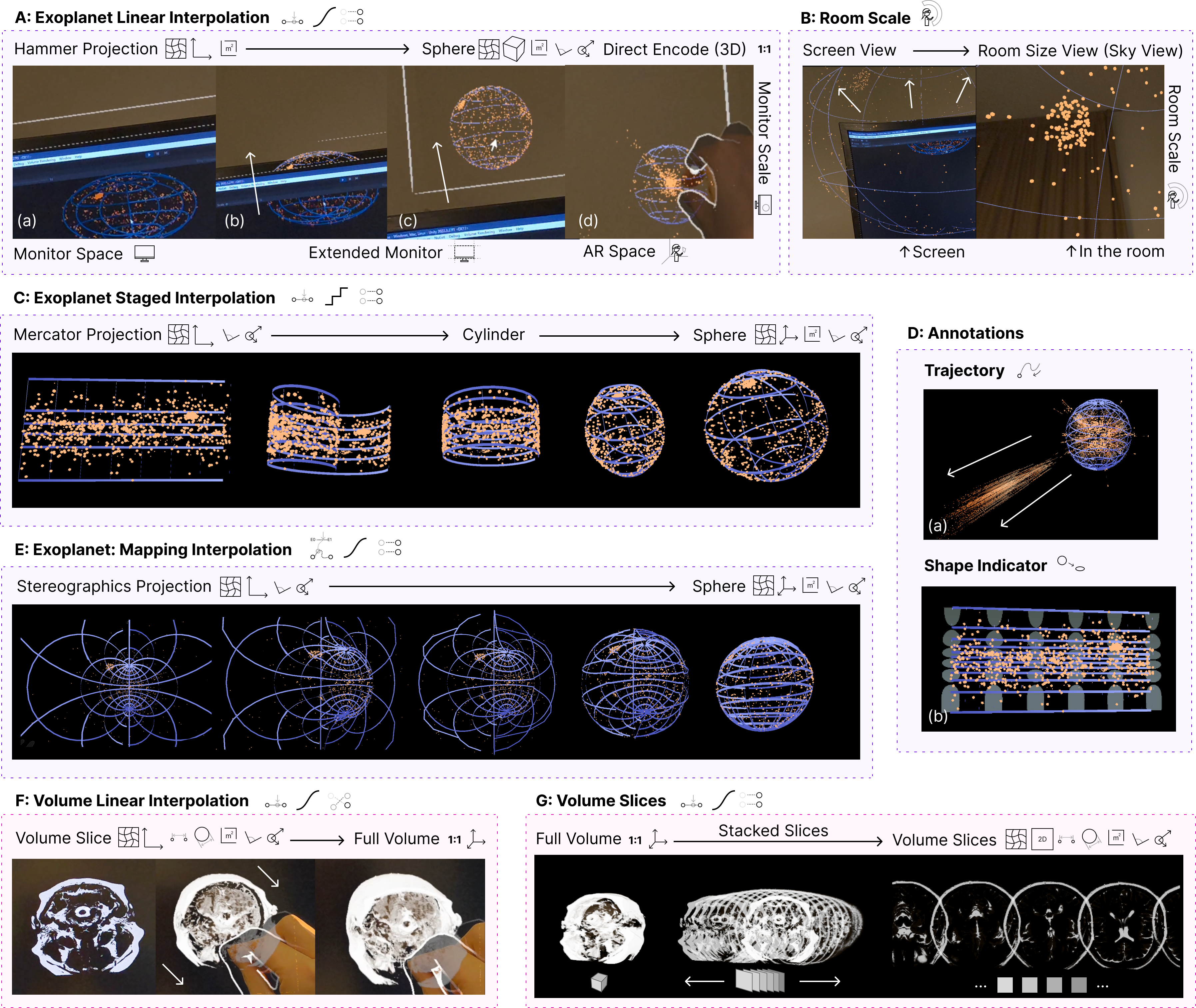}\vspace{-2pt}
  \caption{\label{fig:case_study_figures}%
	\changed{Illustration of a series of transition examples from the exoplanet dataset (A--E) and the brain MRI dataset (F, G). In the figure we also show an example of a room sized representation (B) and of annotations to highlight distortion (D).}}\vspace{-1pt}
\end{figure*}

\textbf{Variation of design choices.}
To map the data to the actual visualization, it is essential to define its appearance in 3D (AR) and 2D (screen) spaces.
\hyperref[fig:stellar_encoding]{Figures~}\ref{fig:stellar_encoding} and \ref{fig:case_study_figures} illustrate some of the visual mappings of the dataset we implemented as examples. 

In 3D, all data points' positional information can be \textbf{directly} mapped to the AR space with a \textbf{monitor scale} (\cref{fig:case_study_figures}(A; d) and \cref{fig:stellar_encoding}(a))---essentially an \emph{outside-in view}. 
\cref{fig:case_study_figures}(B), in contrast, shows a \textbf{room size} view that simulates the sky---an \emph{inside-out view} as if we would look outward from Earth.
Alternatively, the data can be represented as a sphere, sacrificing distance information (\cref{fig:case_study_figures}(A; c) and \cref{fig:stellar_encoding}(b)). 
This approach can serve as a reference for visualizing the stars in the sky as we perceive them from Earth, again as an \emph{outside-in view}. 
Such \textbf{partially} mapped positional data involves discarding some \textbf{distance} and \textbf{length} information during the mapping transition process. 

When projecting the spherical view onto 2D to create an overview, distortion is inevitable. It is thus necessary to consider the retained spatial information after morphs. 
The 2D mappings in \cref{fig:case_study_figures}(A) and \cref{fig:stellar_encoding}(d) use the Hammer projection that retains the \textbf{area} information. 
As another option, the interpolations in \cref{fig:case_study_figures}(B) and (C) use Mercator and stereographic projections (also shown in \cref{fig:stellar_encoding}(c) and (e)). These two projections maintain \textbf{degree} and \textbf{direction}, a characteristic which is also called conformal mapping.

Even though choosing a given projection technique can selectively minimize the introduced distortion for a given application case, additional annotations can still be useful to help users identify this remaining distortion.
\cref{fig:case_study_figures}(D) illustrates a selection of common annotation examples that mitigate distortion issues in cartographic map projections and beyond. 
A \textbf{trajectory}, for instance, is frequently used to track the movement of single data points between two states to visualize the dynamics of the transition procedure. 
In addition, \textbf{coordinates} or \textbf{grids}, such as the Wulff net (also in \cref{fig:case_study_figures}(D; a)), are commonly applied to visualize the space distortion. 
The Tissot indicator~\cite{Karen2001Symbolization} (\cref{fig:case_study_figures}(D; b)) is also a widely used \textbf{shape indicator} to further visualize the distortion of area and anisotropy in a map projection. 

\changed{After finalizing the design decision for the initial and final visual mapping, the next part of designing a transition involves determining the geometric pose.} 
In most scenarios, we apply a \textbf{monitor scale} to the visualization, which offers viewers a convenient and seamless experience for transitioning the dataset using the mouse or gesture between \textbf{AR space} and \textbf{monitor space}. 
The extension of the space of a monitor or display is a common practical approach to enlarge the interactive surface~\cite{reipschlager_designar_2019}, \eg, to improve the reach of mouse interaction. We refer to this setup as \textbf{extended monitor space}. 
This design accommodates the inherently 2D nature of traditional desktop input and screens~\cite{reipschlager_designar_2019}. 
It also avoids the depth issues and the limitation of 2 degrees of freedom (DOF) of input~\cite{Zhou2022DepthMouse}. 

\textbf{Transition.}
\changed{Based on the initial and target visualization of a transition as just defined, we can now construct several morphs between different design choices between two visualizations, for which \cref{fig:case_study_figures}(A) illustrates a common example in our AR-desk\-top environment (also schematically in \autoref{fig:teaser}(A)).} Here, the dataset is initially mapped with the Hammer projection and embedded on the physical monitor. We then drag the visualization into 3D space and transition it to 3D spherical coordinates mapping. 
\changed{With mouse input, the visualization is dragged outside the \textbf{monitor space} and, during this transition, shown in (\cref{fig:case_study_figures}(A; a--c)), simultaneously a transition and a morph happens here: The transition between design choice \textbf{monitor space}, \textbf{extended monitor space} and \textbf{AR space} as well as the morph between Hammer projection (keep \textbf{area} attribute), sphere projection (keep most geometry property except \textbf{distance}), and a \textbf{direct} mapping.
For the animation between the different visualizations, we use \textbf{continuous} in \textbf{interpolation states}, \textbf{parallel} in \textbf{animation sequence}, and \textbf{interpolate geometry} in \textbf{interpolation methods}. This particular transition choice provides users with an uncomplicated solution for a smooth transition experience.}

\changed{Further, \cref{fig:case_study_figures}(C) and (E) illustrate another variation of the transition design. \cref{fig:case_study_figures}(C) uses a \textbf{stepped} transition for the morph between Mercator projection and a 3D sphere to provide viewers with an intuitive explanation by first rolling the map to form a cylinder and then morphing this representation into a sphere.}
\cref{fig:case_study_figures}(E) interpolates between the mapping of stereographic projection and spherical coordinates, which both keep invariant degree information. The interpolation thus also keeps the degree information during the transition and makes the whole transition conformal. The stepped character of this transition separates the whole transition into several semantical intermediate visualizations, in contrast to the continuous representation of the whole single transition. By interpolating the visual mapping rather than directly interpolating the geometric positions, more visual characteristics are retained, thus effectively minimizing extreme distortions during the transition.

\subsection{Brain MRI data}

%Background
\changed{The second data type in our case study is 3D volumetric data: data sampled on regular lattices and often displayed with direct volume rendering. Here, we use MRI data, which is widely used to analyze the internal structure of the human body in medicine~\cite{schultz_topological_2007} (other examples would be CT scans, fluid simulations, etc.).}
\changed{We specifically use the IEEE VIS 2010 contest dataset to demonstrate a volumetric data transition, as schematically shown in \autoref{fig:teaser}(C).}

\changed{We start with a common visualization mapping to explore the volume data's inner structure by utilizing a \textbf{partial} transformation, which shows individual slices.} \changed{\cref{fig:case_study_figures}(F) shows the process of transitioning from a 2D slice (\textbf{partial} transformation) to a \textbf{direct} (volume) representation, in which the slices are sequentially added as the user is `pulling the data out of the screen.'
Adding slices one by one, represents a \textbf{staged} animation sequence design choice.} 
During this transition, we can still observe the intermediate visualization, where the visualization presents a subset of the original visualization---maintaining both its integrity and semantic meaning without any distortion of the spatial information.
\changed{Hence, conserving \textbf{distance}, \textbf{length}, \textbf{angle}, \textbf{area}, and \textbf{direction}.} 
\changed{This example demonstrates that the data's spatial information is not manipulated---we merely show a continuously changing subset, which meaningfully retains a part of the original data and can still be treated as an integrated dataset for mapping designs.}
We can also transition the volume representation to a set of slice-based 2D representations as we show in \cref{fig:case_study_figures}(G), where each slice is one that was originally recorded in the MRI data. Here, we break the \textbf{topology} of the dataset and remove part of its global structure by animating the slices so that they end up being shown side-by-side as traditionally on a surgeon's light box. 
\changed{The slices are fanned out along the x-axis, utilizing the \textbf{interpolate geometry} interpolation method and \textbf{parallel} animation sequence, while keeping the currently selected slice (on the desktop) in the center.
This side-by-side representation does not introduce distortion within the slices, but does, for example, not conserve \textbf{distance} between the slices.}
For volumetric data, of course, it would also be feasible to extract isosurfaces and project them, similar to the mappings we used for the exoplanet dataset, through various design choices.

\subsection{Molecules}

\begin{figure*}
    \centering
    \includegraphics[width=.75\linewidth]{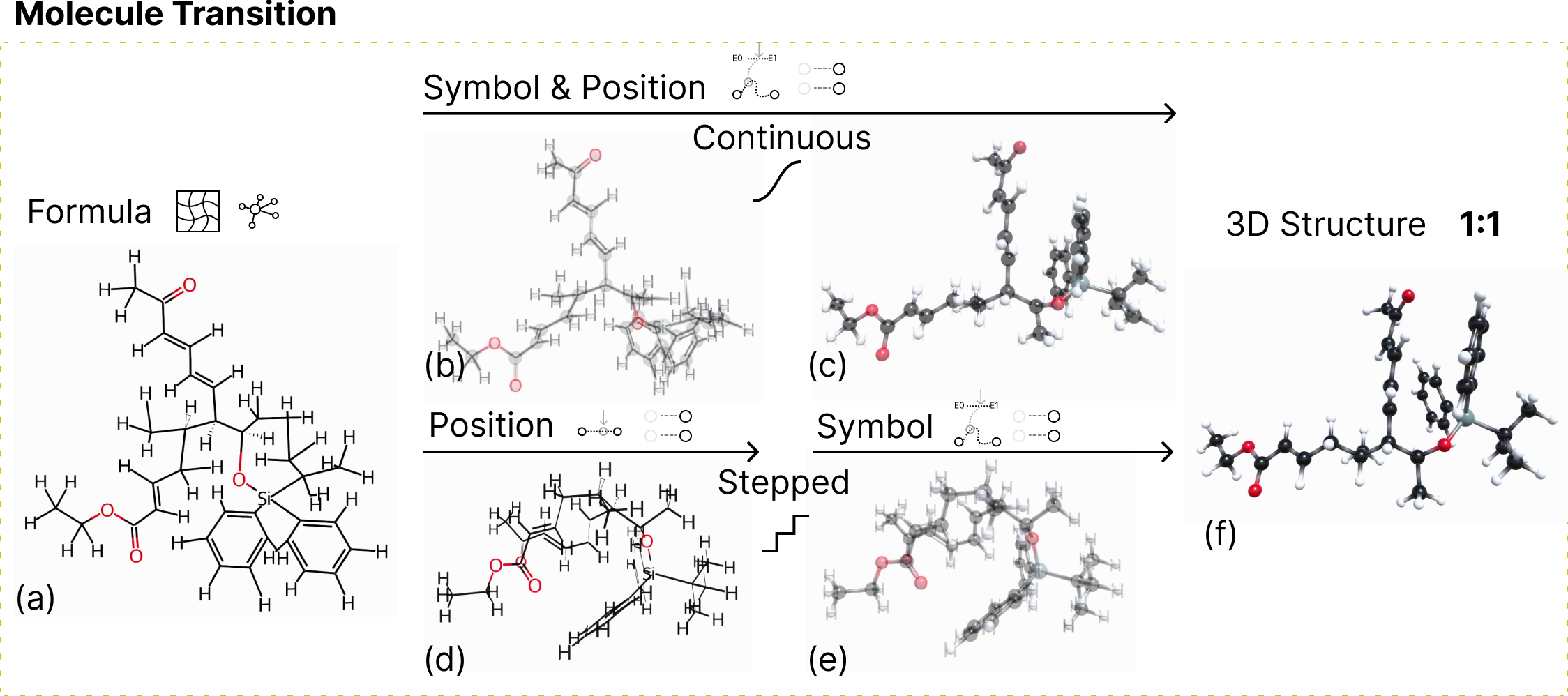}\vspace{-5pt}%
    \caption{Case study on molecular data.
    The atoms' positions are interpolated linearly between the 2D formula and the 3D structure.
    Upper animation path: \textbf{continuous} interpolation; lower animation path: \textbf{stepped} interpolation between the two visualizations.}\vspace{-4pt}%
    \label{fig:molecules}%
\end{figure*}

Finally, we present a use case of structured 3D data in the form of molecular diagrams. 
Graphical representations of molecules have a longstanding tradition in chemistry and biology, as clearly, not only the composition but also the relative spatial arrangement of the atoms determines the properties of a chemical compound. Pertinent to our discussion of the design space, the structure of molecules is not only expressed in 3D but also frequently and uniquely in 2D. 
In contrast to the first case study, here, the transformations are not described by a formula but rather by an algorithm that implements a defined set of rules~\cite{brecher_graphical_2008}.

Traditional representations are known as structural or skeletal formulas (\eg, \cref{fig:molecules}(a)). 
They express which atoms are chemically bound to which other ones and may contain further qualifiers. 
Connecting lines often express the bond type (single, double) and may be related to details of the electron structure (in Lewis formulas, each line represents an electron pair). 
Although generally used as a 2D representation, they can qualitatively express spatial information via, \eg, wedges instead of lines. 
Traditional structural formulas thus primarily express molecule topology, but also provide---for the experienced chemist---a wealth of information about possible reactivity and properties. 
Naturally, a classical 2D projection of the 3D assembly onto a 2D screen cannot represent all the subtle details of a molecular conformation, and the spatial extent of certain parts of the molecule may not be obvious. Yet, this information is important for the stereochemical selectivity of reactions and the overall reactivity, so 3D views are used as well (\eg, \cref{fig:molecules}(f)). 
Moreover, for larger molecules, the mapping of 2D structural formulas and 3D spatial models may not be obvious---even for experienced chemists. 

The molecule in our example (\cref{fig:molecules}) contains two double bonds in one of the side chains and another double bond in the other side chain. 
There is thus the possibility of an intramolecular reaction, which is quite obvious for chemists due to their familiarity with the representation. 
A reaction can only take place, however, if the spatial arrangement allows it, which is not visible in the 2D representation, and we thus transition the structural formula to its known 3D structure (\cref{fig:molecules}(f)). 
For this purpose, we apply spatial coordination during the morph from structure formula to a 3D ball-and-stick representation, \ie, we move the letter and line elements to their correct 3D positions as we transition. 
Yet, such a \textbf{continuous} interpolation (\cref{fig:molecules}(b, c)), when done on a desktop PC only, still requires interactive camera movement to fully grasp the alignment of the two side chains. 
When the morph is supported by a transition into the 3D/AR environment, in contrast, the chemist can fully concentrate on tracking the location of the double bonds. 
Their visibility and tracking can further be enhanced by a \textbf{stepped} interpolation (\cref{fig:molecules}(d, e)). 
Here, in the first step, we morph the atoms of the structure formula to their respective 3D positions and, afterward, smoothly blend into the 3D balls-and-sticks model. 
In both cases, we use a representation at \textbf{monitor scale}, but the \textbf{room scale} can be beneficial to investigate, \eg, datasets of crystals. 
Here, we chose the \textbf{position space} as \textbf{AR} because problems in chemistry often require a certain view angle on the molecules for occlusion-free vision on the region of interest, which requires interaction to rotate the molecule. 
The 3D manipulation using hand gestures is easier when done close to the user than next to the monitor. 

A \textbf{staged} animation sequence is a possible alternative to enhance the spatial sense-making during the morph. 
For example, each side chain of the presented molecule could be animated in a staged fashion, which allows the chemist to concentrate only on one part of the molecule at a time. 
The trade-off here is that this would create several intermediate visualizations that are chemically ``incorrect''. 
However, adding an animated \textbf{visual encoding} that reduces the opacity of the side chains that are currently not animated or in the final visualization could mitigate this effect. 
\textbf{Annotations} are either not possible to implement statically, like \textbf{distortion}, or would only clutter the visualization since the 3D visualization of molecular structures is already quite complex.

\section{Discussion}
\label{sec:discussion}

While these case studies demonstrate the potential application of our design space, we also want to raise several points that affect its application and use in the future as well as its potential extension. 

\textbf{Notion of scale.} In most cases, 3D spatial data inherently provides a literal scale of the data. 
Whether or not it is feasible to present the data on the literal scale, especially in cases where the literal scale is between monitor and room scale, is up for further investigation. 
In the examples we presented in our case studies, only the MRI dataset can be presented at its literal scale, which is, depending on the required analysis process, probably not the preferred choice by domain scientists. 
Therefore, the question of how to scale the data during the transition between environments strongly depends on the task at hand. 
Since the scale is difficult to generalize, we believe that giving users control over the size is the best solution. 
 
\textbf{Composed visualizations.} Visualizations of 3D spatial data that are combined with 2D annotations or visualizations \cite{hong_survey_2024-1} pose a whole new set of challenges in the context of AR-desktop hybrid environments. 
In a transition, for example, the 2D parts of the visualization could stay on the desktop PC, and only (annotation) lines could be provided to link between the representations and environments~\cite{schwajda_transforming_2023}. 
But this process could also include a different set of transformation operations to reach a different final state of the 2D visualization. 
Therefore, we could separate the 2D and 3D parts of the visualization into two categories and regard the information separately during the animated transition. 
Our design space provides some options to separately process the visualizations, but would require further dimensions to provide comprehensive options for this kind of visualization.

\textbf{Intermediate states.} 
%The intermediate states that result from \textbf{staged} animation sequences and \textbf{stepped} interpolation distinguish animated transitions we employ from simple animations. 
%They provide a meaningful transition sequence for the users rather than simply using a slightly more engaging process \cite{TVERSKY2002247}, especially when combined with an \textbf{interpolation of mapping}. 
\changed{A transition design that emphasizes intermediate states by, \eg, using \textbf{interpolate mapping} together with a \textbf{staged} animation sequence or \textbf{stepped} interpolation states, provides valuable additional information that supports users in mentally connecting two representations.}
In our molecule use case, \eg, the intermediate state, in which the structure formula is arranged the same way as the balls-and-sticks visualization, provides a way to connect textbook knowledge with the spatial arrangement of the molecule. 
Some spatial datasets lend themselves well to intermediate states, but in general, they are quite difficult to create automatically, even with the utilization of our design space.
\changed{Also, providing full control over the animation timeline (\eg, a scrollbar of a video) can be a powerful tool for data exploration using animated transitions.}

\textbf{Interaction for transitions.} Our design space focuses on integrating spatial information and on the relationship between visualization and the display environment. 
\changed{While the original data contains the spatial information that is potentially sampled from reality, presenting the data within the AR-desktop space may also semantically assign real-world knowledge to the visualization or interaction.}
The area surrounding the monitor, for instance, can imply different types of transitions based on the direction relative to the monitor~\cite{reipschlager_designar_2019}. 
These spatial semantics can be stacked or juxtaposed to create complex layouts for more sophisticated designs. 
In addition, the position in the AR-desktop environment can be represented by three variables. 
Based on the positioning in the immersive environment, it is thus feasible to assign up to three parallel input variables. 
Each variable can correspond to a distinct state and can support multiple state changes simultaneously. 
The monitor's physical appearance is also usually rectangular, which can be used as a reference frame. 
As we focused on visual appearance in our work, however, we did not include such considerations as dedicated design dimensions. 

\textbf{Multiple datasets.} While our case studies only come from three domains (yet with a representative range of different types of data), it is important to consider the broader context of data analytics systems. 
Each domain typically has unique spatial data types and specific tasks that are not easily generalized across all cases. 
It is also common for visual analytics systems to display multiple or combined visualizations simultaneously \cite{kreiser_survey_2018}. 
A user may have different visualizations of the same dataset; each presenting distinct aspects of the data. 
We hypothesize that this concept can be applied to lay out multiple visualizations and, thus, open the possibility to transition between them. 
In addition, presenting the same information through different visualizations may help us reduce bias and enhance the user's overall understanding of the data.

\textbf{Animation.} We believe that the dimensions of our design space that influence the sequence of elements or glyphs have far more impact on a user's experience than animation curves or staggered animations. 
Nonetheless, being able to swap animation curves is a possible additional design dimension. 
Our approach inherently lets the user control animations while transitioning a visualization. 
\changed{We believe, however, that further control, like moving the object in a direction that is orthogonal to the direction that triggers the transition, could possibly be used, or drag-and-drop operations could initiate dedicated animation sequences.} 

\textbf{Generalizability.} Describing transitions using our proposed design space is highly domain-specific.
In general, it may be difficult to translate concepts that work on one domain's dataset to another domain. 
This fact also implies that some of our proposed design dimensions can be extended by considering further domains and datasets, especially for our annotation design dimensions. 
\changed{While our work focuses specifically on the AR-desktop environment, the visualization transition design within our chosen environment has the potential to extend to other cross-de\-vice settings, such as those involving tablets and AR headsets. When portable or multiple devices are involved, the geometric pose design space may further consider the spatial relationships among devices, the physical environment, and the visualizations. In addition to cross-de\-vice scenarios, transitions across different levels of vir\-tu\-a\-li\-ty---such as between VR, AR, and physical-world monitors \cite{frohler_survey_2022, aigner2023Cardiac}---al\-so require considerations when designing transitions between different actualities.}

\textbf{Hybrid input modalities.} \changed{For our prototypes, we developed a simple framework that focuses on AR-desktop environments. 
For a proper generalization of interaction across multiple devices~\cite{Horak_2019_Vistribute}, there remains a need for a more universal framework to integrate different input devices while supporting smooth input transitions and seamless manipulation of virtual objects.}
Though existing studies\cite{hubenschmid_relive_2022,schroder_2023_collaborating,Rau:2025:TDR} debate whether users like to change input modalities,
we believe that the traditional mouse and keyboard input will not be replaced anytime soon for desktop PCs \cite{wang_towards_2020,pavanatto_we_2021}. 
\changed{For hybrid systems, however, developers should consider input modalities that can be used on all incorporated devices without the requirement of switching, and which then could be used to control the transitions we describe in our design space.} 

\changed{\textbf{User feedback.} In this work, we deliberately did not conduct an empirical experiment to validate our work, as our focus is the design space rather than evaluating the usability of a specific implemented instance of the interaction design or of given use cases. Our goal is to empower designers to describe, generate, and evaluate designs \cite{BeaudouinLafon:2021:GTI}. Given the domain-specificity of spatial dataset visualization tasks, a user study would have only offered limited general insight into the design space. Instead, we validated our approach through our diverse use case demonstrations in \autoref{sec:case_studies_sec}. Moreover, prior studies have already examined the general effects of AR-desktop hybrid systems and demonstrated the user engagement in transferring virtual objects between screen and AR space \cite{cools_comparison_2025,liao2025seamlessvr,Rau:2025:TDR}.}%\vspace{-3pt}

\section{Conclusion}
With our design space for transitions of spatial data between planar 2D displays and stereoscopic AR environments, we extended past work that looked at such transitions for non-spatial data \cite{lee_design_2022}. 
With our design space, we demonstrate how to make use of and cater to the unique spatial properties of data that arise in many scientific application domains such as medicine, astronomy, chemistry, physics, etc., for which we are not at liberty to use the third dimension in AR space to facilitate the transition between the different actualities. 
Instead, in our design space, we embrace this constraint and demonstrate how we can make use of different spatial mappings, annotations, and dedicated animation sequences to allow viewers to mentally follow the changing representation of the data as it transitions between the spaces. 
As we noted in our discussion, we do not make the claim that our design space would be complete. 
Domain-specific constraints and practices are likely to give rise to additional possibilities for designing transitions that, in turn, may then be applicable to other use cases. 
Nonetheless, we have laid the foundation for the discussion of how to design effective AR-desktop environments for the analysis of spatial data.%\vspace{-3pt}

% The MRI Brain dataset is courtesy of Prof. B. Terwey, Klinikum Mitte, Bremen, Germany, from the IEEE Visualization Contest 2010.

\section*{Acknowledgments}
We thank Ambre Assor for her feedback on our manuscipt. %\vspace{-3pt}
This work is partially funded by Deutsche Forschungsgemeinschaft (DFG, German Research Foundation) under Germany's Excellence Strategy---EXC 2075--390740016. We also acknowledge the support by the Stuttgart Center for Simulation Science (SimTech).

\section*{Supplemental Material Pointers}
We share our additional material (author version of the paper, self-created figures, video figure) at \osfrepo. We will also make the source code available on GitHub upon publication of the paper.
% All source code can be found at \href{https://github.com/tingying-he/design-characterization-for-black-and-white-textures-in-visualization}{\texttt{github\discretionary{}{.}{.}com\discretionary{/}{}{/}tingying\discretionary{}{-}{-}he\discretionary{/}{}{/}design\discretionary{}{-}{-}characterization\discretionary{}{-}{-}for\discretionary{}{-}{-}black\discretionary{}{-}{-}and\discretionary{}{-}{-}white\discretionary{}{-}{-}textures\discretionary{}{-}{-}in\discretionary{}{-}{-}visualization}}.}

\section*{Images/figures license/copyright}
We as authors state that all of our figures in this article are and remain under our own personal copyright, with the permission to be used here. We also make them available under the \href{https://creativecommons.org/licenses/by/4.0/}{Creative Commons At\-tri\-bu\-tion 4.0 International (\ccLogo\,\ccAttribution\ \mbox{CC BY 4.0})} license and share them at \osfrepo.

%-------------------------------------------------------------------------
% bibtex
\bibliographystyle{eg-alpha-doi} 
\bibliography{abbreviations,references}       

% biblatex with biber
% \printbibliography                

\end{document}